\documentclass[fleqn,10pt]{wlscirep}

\usepackage{amsmath}
\usepackage{amssymb}
\usepackage{graphicx}

\usepackage[FIGTOPCAP,tight,raggedright,nooneline,bf,sf]{subfigure}

\graphicspath{{images/}{plots/}}

\title{Phase Transitions of the Typical Algorithmic Complexity
    of the Random Satisfiability Problem Studied with Linear Programming}
\author[1,*]{Hendrik Schawe}
\author[1]{Roman Bleim}
\author[1,\&]{Alexander K. Hartmann}

\affil[1]{Institut f\"ur Physik, Universit\"at Oldenburg, 26111 Oldenburg, Germany}
\affil[*]{hendrik.schawe@uni-oldenburg.de}
\affil[\&]{a.hartmann@uni-oldenburg.de}

\begin{abstract}
    Here we study the NP-complete
    Satisfiability problem for $N$ Boolean variables, in particular
    $K$-SAT, for which the Boolean
    formula has the conjunctive normal form with $M$ clauses and
    $K<N$ possibly negated  variables per clause.
    Although the worst-case complexity of NP-complete problems
    is conjectured to be exponential, there exist
    parametrized random ensembles of problems where solutions can
    \emph{typically} be found in polynomial time for suitable ranges
    of the parameter.
    In fact, random $K$-SAT, with $\alpha=M/N $ as control parameter,
    can be solved quickly for small enough values of $\alpha$. It shows a phase
    transition between a satisfiable phase and an unsatisfiable
    phase. For branch and bound algorithms, which operate
    in the space of feasible Boolean configurations, the empirically
    hardest problems are located only close to this phase transition.
    Here we study $K$-SAT ($K=3,4$) and the related optimization
    problem MAX-SAT by a \emph{linear programming approach},
    which is widely used for practical problems
    and allows for polynomial run time.  In contrast
    to branch and bound it operates outside the
    space of feasible configurations. On the other hand, finding a solution
    within polynomial time is not guaranteed.
    We investigated several variants
    like including artificial objective functions, so called
    \emph{cutting-plane approaches}, and a mapping to the NP-complete vertex-cover
    problem. We observed several \emph{easy-hard transitions}, from
    where the problems are typically solvable (in polynomial time) using
    the given algorithms, respectively, to where
    they are not solvable in polynomial time.
    For the related vertex-cover problem on random graphs
    these easy-hard transitions can be identified with structural properties
    of the graphs, like percolation transitions.
    For the present random $K$-SAT problem we have investigated
    numerous structural properties also exhibiting clear transitions,
    but they appear not be correlated to the here observed easy-hard
    transitions. This renders the behaviour of random $K$-SAT more
    complex than, e.g., the vertex-cover problem.
\end{abstract}

\begin{document}
    \flushbottom
    \maketitle
    \thispagestyle{empty}

\section*{Introduction}
    The \emph{Satisfiability problem} (SAT) \cite{garey1979}
    is to decide whether some Boolean formula is satisfiable or not, i.e.,
    whether for a given Boolean formula, there is an \emph{assignment} of the
    variables such that the formula evaluates to ``true''.
    SAT is the most-prominent NP-hard (nondeterministic-polynomial) problem.
    NP-hard problems \cite{garey1979,papadimitriou1998}
    are fundamental to computational complexity, since NP-hard means that all
    problems in NP can be mapped in polynomial time to any of them.
    Despite much effort no algorithm has been found so far, which is able to
    solve any NP-hard problem in the \emph{worst case} in polynomial time,
    leading to the famous \emph{P-NP problem}. Thus, all NP-hard problems
    are considered so far being \emph{hard}.
    SAT is in particular \emph{NP-complete} \cite{cook1971complexity}, which
    means that given a solution it can be verified in polynomial time and is
    therefore member of the class of NP problems.
    Thus, if one day someone found a fast algorithm for SAT (which most
    researcher believe will never happen), one could efficiently solve \emph{all}
    problems in NP, especially also all NP-complete problems. Anyway, also to
    advance knowledge helping to prove that actually no fast algorithm exist,
    one wants to understand the reason for the apparent computational hardness.
    One somehow empirical approach is to analyse actual (relatively) hard
    instances of problems. This has attracted  much interest in statistical
    physics \cite{phase-transitions2005,mezard2009,moore2011} and correspondingly
    \emph{random ensembles} of suitably parametrized problem instances
    and their \emph{typical hardness} have been investigated.

    This has been done also for SAT, in particular for
    a subclass defined as follows:
    All Boolean formulas can be expressed in \emph{conjunctive normal form}
    (CNF) which is a disjunction of \emph{clauses}, each being a conjunction
    of variables or negated variables.
    Therefore $K$-SAT, which is a Boolean formula in CNF with $K$ distinct
    variables per clause, is a commonly scrutinized version
    of the satisfiability problem. The most prominent random ensemble for SAT
    is the random $K$-SAT,
    where for a set of $N$ variables $M$ clauses are generated randomly.
    Each clause contains $K$ distinct variables which are chosen randomly,
    and each variable appears negated with probability $0.5$.
    Interestingly, this problem shows a \emph{phase transition} at some
    critical value $\alpha_{\mathrm{s}}$ of the density
    $\alpha=M/N$ \cite{cheeseman1991really}. For
    large problems at $\alpha < \alpha_\mathrm{s}$
    almost all problems are satisfiable (also denoted as SAT),
    above $\alpha_\mathrm{s}$ almost all realizations
    are unsatisfiable (UNSAT).
    The occurrence of similar phase transitions has been observed
    frequently for other random ensembles of NP-complete
    problems \cite{cheeseman1991really,phase-transitions2005,mezard2009,moore2011},
    for which 3-SAT is the prime example \cite{cook1971complexity,karp1972reducibility}.
    This incited strong interest in the $K$-SAT problem \cite{mezard2002,biroli2002phase,krzakala2007gibbs,cocco2001}
    and many other NP-complete problems \cite{martin2001statistical,gent1996tsp,Zdeborova2007phase,vertexCover2,vertexCover1,cover-time2001}
    among physicists.

    For the SAT-UNSAT transition it was found that the hardest realizations
    are located near this $\alpha_\mathrm{s}$. This can be roughly
    understood because
    it is trivial to find some solution for much lower values of $\alpha$
    and relatively easy to prove a realization unsatisfiable at high values of $\alpha$.

    While this SAT-UNSAT transition is certainly the most scrutinized
    in the $K$-SAT problem, there exist more transitions. For example 3-SAT, where
    SAT-UNSAT occurs at $\alpha_{\mathrm{s}}\approx 4.26$ \cite{mezard2002},
    shows a transition to chaotic behaviour at $\alpha_{\chi} \approx 3.28$
    \cite{varga2016order}, i.e., using a continuous time deterministic
    solver \cite{ercsey2011optimization} the trajectory will find the
    solution if one exists, but it will show chaotic transient behaviour
    above this threshold resulting in increasing escape rates from
    attractors.  This leads to a higher computational cost and can therefore
    be used as a measure of hardness.  Furthermore, there exists a
    clustering transition at $\alpha_{\mathrm{c}} \approx 3.86$
    \cite{krzakala2007gibbs,mann2010numerical}.  This means that here the
    organization of the space of the exponentially many degenerate
    solutions changes from one big cluster ($\alpha<\alpha_{\mathrm{c}}$)
    of solutions which are connected in assignment space to a solution
    space ($\alpha>\alpha_{\mathrm{c}}$) which is fragmented into many
    non-connected smaller clusters, one says replica symmetry is broken above
    this threshold.

    Usually algorithms like the \emph{branch and bound approach},
    \emph{stochastic search} or \emph{message passing}
    are used in the statistical-mechanics literature.
    These algorithms operate in the space of feasible assignments
    and approach the optimum solution from above. Here ``optimum'' means that
    the number of unsatisfied clauses is  minimized (in the sense
    of MAX-SAT), i.e., eventually becomes
    zero if a satisfying assignment is found. Thus for general minimization
    problems these algorithms
    yield upper bounds until the true minimum solution is found.
    As empirically studying the computational hardness always
    tells something about a problem in conjunction with a specific algorithm,
    it is desirable to investigate different algorithms, in particular
    approaches which differ fundamentally.
    The operations-research literature often uses
    \emph{linear programming} (LP) \cite{papadimitriou,cormen2009introduction},
    which operates for combinatorial problems
    outside the space of feasible solutions, i.e., fundamentally different
    from the above mentioned algorithms. Suitable enhanced (see below)
    LP techniques are often used
    for real-world applications since they are versatile and efficient,
    which means they run typically in polynomial time.
    For combinatorial problems, e.g., NP-hard
    optimization problems, the application of pure
    LP yields solutions which are not necessarily
    feasible, which here means non-integer-valued assignments to the variables,
    but which establish a lower bound on the objective.
    Thus LP somehow approaches for minimization problems the true
    feasible and optimum solution from below (In this sense the analog solver of \cite{varga2016order,ercsey2011optimization}
    also operates outside the feasible region). Nevertheless, a key observation is
    that whenever LP gives a feasible solution, it must be the true optimum solution
    of the combinatorial problem.
    Because of their complementary properties combinations of LP with other
    approaches yield powerful methods,
    such as \emph{branch and cut} algorithms \cite{padberg1991branch}.
    Here \emph{cutting planes} (CPs) are employed, which are additional
    inequalities which decrease LP solution space to help finding
    feasible solutions.

    Although LP, in combination with other techniques, is an
    (efficient) optimization algorithm for practical
    optimization problems, there are only few studies aiming
    at understanding the nature of hardness and detecting phase
    transitions. To our knowledge,
    studies of the behaviour of LP have only been conducted for
    the vertex cover (VC) \cite{dewenter2012phase} and the
    travelling salesperson problem (TSP) \cite{schawe2016phase}. For
    these problems there
    exist regions in parameter space, where feasible and optimal
    solutions can be found in polynomial time.
    For VC, the LP approach yielded solutions up to the
    percolation transition of the underlying graph ensemble. Therefore,
    the problem is \emph{easy} with respect to LP up to the
    percolation transition, and \emph{hard} beyond.
    For LP improved with cutting planes, another easy-hard transitions
    occurs at the point of the onset of replica symmetry
    breaking \cite{vertexCover2}, which here corresponds to the
    clustering transition of SAT. Note that
    this coincides with the percolation threshold for the \emph{leaf-removal
    core} \cite{bauer2001core}.
    This is the point where one would not reasonably expect easy instances
    anymore.
    For TSP the easy-hard transitions detected by the LP approach coincided
    with structural changes of the optimal tour which can be intuitively
    understood as increases in hardness. Since for TSP many more cutting planes
    exist, which are not yet tested, it is conceivable to use this technique
    to find more and more easy-hard transitions this way and understand them,
    leading to deeper insight into the problem.

    Since $K$-SAT is the archetypal NP-complete problem, we wanted to extend
    those promising results of the mentioned previous studies.
    Although $K$-SAT is by definition a decision problem and
    not an optimization problem, as we will
    show below, we found also some ``easy-hard'' transitions. Nevertheless, these
    transitions occur at much lower values of the parameter $\alpha$ than
    the clustering transition, in contrast to VC on random graphs.
    This result could be related to the fact that $K$-SAT shows
    anyway a richer behaviour \cite{krzakala2007gibbs}, i.e.,
    several different types of transition, in contrast
    to VC on random graphs. Anyway, we cannot explain this result in the moment,
    and think it will motivate further studies which aim at the origin of the
    different behaviour.

    Note that this study does not aim to present faster methods to solve the
    SAT problem, but rather tries to study it in a fundamental sense,
    aiming at the question ``What makes a problem hard''?
    We pursue this by  applying an  approach which is widely-used for many
    practical combinatorial optimization problems, but less-often
    studied when considering
    $K$-SAT, in particular in the computational complexity and
    statistical-physics communities. In fact,
    we will show that the easy-hard transitions with respect to the used
    LP algorithms happen at rather low values of $\alpha$. Thus, other
    more specialized algorithms are preferable for practical $K$-SAT
    solving.

\section*{Models and Methods}
\label{sec:methods}

\subsection*{$K$-SAT}
    A realization of $K$-SAT consists of a Boolean formula over $N$
    variables $x_i$ ($i=1, \ldots, N$). The formula is
    in conjunctive normal form, i.e., it is a conjunction of $M$ clauses
    $c_j$ ($j=1, \ldots, M$), where every clause is a disjunction of $K$ literals
    $l_{kj}$ ($k=1, \ldots, K$). A literal is a
    variable $x_i$ or its negation $\overline{x}_i$.
    In each clause, each variable may appear only once.
    As an example for $N=4$, $M=2$ and $K=3$ take
    \begin{align}
        \label{eq:sat_example}
        (\overline{x}_1 \vee x_2 \vee \overline{x}_3) \wedge (x_1 \vee x_3 \vee \overline{x}_4).
    \end{align}
    This example is solvable with, e.g., $x_1 = \text{``true''} = 1$ and
    $x_3 = \text{``false''} = 0$,
    and arbitrary assignments for the other variables. Note that
    each clause is satisfiable by $2^K-1$ out of $2^K$ possible assignments to
    the variables. Thus, each clause restricts the space of satisfiable assignments a bit.
    Clearly, with more clauses per variable, i.e., a higher amount of
    constraints, it is more probable that a random formula is unsatisfiable.
    As mentioned before, for random 3-SAT with $N \to \infty$ there is a critical
    density $\alpha_\mathrm{s} = M/N \approx 4.26$ \cite{mezard2002}
    at which a phase transition from satisfiable to unsatisfiable (SAT-UNSAT) happens.

\subsection*{Linear Programming}

    A linear program (LP) is an optimization problem, which can be expressed by
    a set of linear constraints and a linear objective function, which should
    be optimized. There are fast (polynomial-time) algorithms
    to solve a linear program, e.g., the ellipsoid method \cite{khachiyan1980polynomial} or interior
    point methods \cite{karmarkar1984new,nesterov1994interior}. However,
    in many sophisticated solvers the simplex algorithm is used, which
    typically terminates quickly for real-world problems, despite its
    exponential worst-case time complexity \cite{papadimitriou,cormen2009introduction}.
    Though, as soon as some variables need to be integer
    valued, this problem gets computationally hard.
    In fact, \emph{integer linear programming} is
    an NP-hard problem \cite{karp1972reducibility}.

    A $K$-SAT realization can be expressed as an integer linear program. Therefore every
    positive literal $x_i$ is expressed as an integer variable $x_i$ and every
    negative literal $\overline{x}_i$ as $(1-x_i)$. Since one or more
    literals of every
    clause $c \in C$ need to be true for a satisfying assignment, the
    corresponding integer linear program contains for each clause the
    constraint that the sum of the expressions for the
    included literals must be greater or equal 1. The example from Eq.~\eqref{eq:sat_example}
    generates following linear inequalities.
    \begin{align*}
        (1-x_1) + x_2 + (1-x_3) \ge 1\\
        x_1 + x_3 + (1-x_4) \ge 1
    \end{align*}

    Since an LP is an optimization problem but SAT is merely a decision problem, we can
    choose an arbitrary objective function for which to optimize.
    The simplest objective function is zero, i.e., no optimization.

    \begin{align*}
        &\text{min.} & 0\\
        &\text{s.t.}  & \sum_{x_i \in c_j} x_{i} + \sum_{\overline{x}_i \in c_j} (1-x_{i}) & \ge 1,       & & \forall 1 \le j \le M\\
        &                  &  x_{i}                                                       & \in \{0,1\}, & & \forall 1 \le i \le N
    \end{align*}

    The last constraint fixes the variables to integer values. We will relax
    this constraint to $x_i \in [0,1]$. This allows us to apply a fast
    LP algorithm to solve the relaxed problem  and introduce a measure of
    hardness for the problem realization. If the LP relaxation yields a solution
    consisting of only integer variables, the solution is obtained by a polynomial
    time method and the corresponding realization is obviously easy to solve.

    A drawback is that additionally to the principal degeneracy of the
    problem, i.e., there are possibly many assignments that satisfy the
    formula, the relaxation leads to a much higher degeneracy. For
    example, the assignment of all $x_{i} = 0.5$ is always a solution
    of the relaxation.
    After we performed some simulations for SAT in this way,
    it became evident that this
    degeneracy is a major problem for this decision problem, which is not
    present in the optimization problems studied with this method \cite{dewenter2012phase,schawe2016phase}.
    This degeneracy leads to different behaviour for slight changes in the
    algorithm. E.g., primal and dual simplex often lead to different behaviour
    such that for many instances one version will result in an integer solution
    while the other does not.
    We observed a similar behaviour when considering
    different pricing strategies or a presolve stage to tighten
    the LP. This analysis would therefore only yield information about
    these technical
    details and not about the fundamental problem of $K$-SAT. For example,
    the presolver of both Gurobi and CPLEX can solve easy instances up to an
    critical $\alpha_{\mathrm{pre}} = 1.640(1)$, which is the same threshold
    up to which the \emph{pure-literal} rule (also called
    \emph{affirmative-negative rule}) which is an integral part of the
    DPLL \cite{davis1960,davis1962} search algorithm, can solve $K$-SAT realizations,
    while without presolve the easy-hard transition occurs at a lower value
    of $\alpha$ --
    dependent on technical details of the method.
    Therefore, we do not present results of LP with zero objective in is study.

    Instead, we will introduce artificial objective functions to reduce the
    degeneracy drastically. Further, the choice of the objective function
    has an influence on the prevalence of integer solutions.
    Though note, only linear objective functions enable the
    efficient linear programming techniques. Therefore,
    non-linear objectives like $\sum_i x_i(1-x_i)$, which
    are minimal if all variables $x_i$ are either $1$ or $0$, are not admissible
    and in fact generally NP-hard \cite{Sahni1974}.

    One simple way to replace the zero objective is maximizing
    the sum over all variables (MV)
    \begin{align*}
        &\text{max.} & \sum_{i=1}^N x_i,
    \end{align*}
    which will on average lift variables which are $0$ in the integer solution
    to larger values like $0.5$ and thus suppresses integer solutions typically.

    Note that this and other additional objective functions have no
    influence on whether a formula is satisfiable or not, they are just meant
    as a tool to reduce the degeneracy of the problem to make it less
    dependent on details of the algorithm and to facilitate finding integer
    solutions. Both works out as we will see below.
    As a third objective we tried maximizing the number of fulfilled
    literals per clause, which we will call
    \emph{Satisfaction Multiplicity Maximization} (SMM).
    This can be achieved with a slightly modified linear program
    by introducing one new variable $z_j$ per clause counting the
    number of fulfilled literals of its clause and maximizing the
    sum over all $z_j$.

    \begin{align*}
        &\text{max.} & \sum_{j=1}^M z_j\\
        &\text{s.t.} & \sum_{x_i \in c_j} x_{i} + \sum_{\overline{x}_i \in c_j} (1-x_{i}) & \ge z_j,     & & \forall 1 \le j \le M\\
        &            &  x_{i}                                                             & \in \{0,1\}, & & \forall 1 \le i \le N\\
        &            & z_j                                                                & \ge 1,       & & \forall 1 \le j \le M
    \end{align*}

    The new kind of constraint ensures that $z_{i} \ge 1$, i.e., that
    every clause contains at least one fulfilled literal, such that the
    solution assignment satisfies the Boolean formula.
    This type of additional optimization is similar to MAX-SAT, where
    one tries to maximize the number of satisfied clauses. For MAX-SAT one
    would instead enforce $0\le z_j\le 1\; \forall j$. We also tried
    this MAX-SAT approach to solve the $K$-SAT decision problem.
    Since this formulation does not mitigate the degeneracy problem, and
    because we did not observe any better
    performance than by using the other approaches, we do not show results
    for this approach here.

    The SMM objective is an example for a linear objective with a slight
    preference for integer valued variables,
    since assignment of all variables of a clause to $1$ or $0$ according to
    their polarity contributes more to the objective function than assignments
    of non-integer values. For example a variable which appears more often
    unnegated will on average be assigned more often to value 1.
    This strongly reduces the degeneracy of the
    solution of the optimization problem, i.e., many of the solutions
    where the majority of variables are non-integer, are not optimal under this
    new objective function.
    Of course this may still yield non-integer values for some variables,
    which occur in conflicting clauses.

    Note that when using LP, finding integer solutions may be facilitated
    in principle by adding cutting planes, which are further
    constraints which are added to the problem during run time dependent on the
    state of the solution process. This allows, e.g, to add
    a selected small number of constraints from a set
    of exponentially many ones,
    for which it would be infeasible to add them all.
    This was previously observed also for ensembles of random
    instances for the vertex
    cover \cite{dewenter2012phase} and the TSP \cite{schawe2016phase}.
    Nevertheless,
    for the present study this yields no improvement, i.e., no additional phase
    transitions could be observed, see below. Thus, we do not describe the
    cutting-plane approach in this section in detail, but rather just list the
    the ones we tried and did not lead to an improvement.

\subsection*{Mapping to Vertex Cover}

    \begin{figure}[htb]
        \center
        \includegraphics[scale=1]{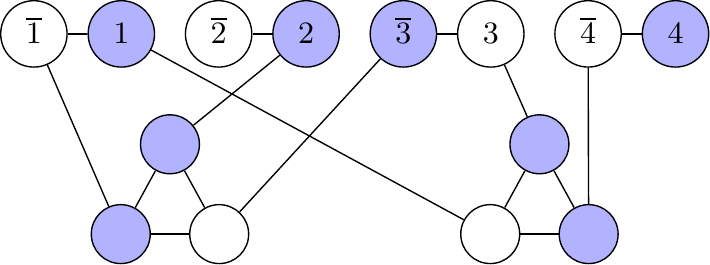}
        \caption{\label{fig:vc:sat}
            The graph for the vertex cover problem which is equivalent
            to  the formula shown in Eq.~\eqref{eq:sat_example}. Shown is a vertex cover of
            size $N+(K-1)M=4+2\times 2=8$, which corresponds to a satisfying assignment
            $x_1=1$, $x_2=1$, $x_3=0$, $x_4=1$.
        }
    \end{figure}

    All NP-complete problems, by definition, can be mapped onto
    each other in polynomial time. Thus, it is reasonable to ask, whether
    transforming SAT instances to instances of another problem
    and applying algorithms specifically suited for the other problem
    changes the performance, as measured by the location of the easy-hard transition.
    Here we used a classical mapping \cite{garey1979}
    of SAT to VC. For each $K$-SAT instance
    an equivalent graph $G=(V,E)$ is constructed in the following way: The
    set $V$ of nodes contains
    one pair of nodes $i, \overline{i}$ for each variable $x_i$ ($i=1,\ldots,N$),
    which represents the variable and its negation. Furthermore, $V$ contains
    one node $(kj)$ for each literal $l_{kj}$ ($k=1,..,K$, $j = 1,..,M$) in each
    clause $c_j$, respectively.
    Therefore $V$ contains $2N+KM$ nodes. For the
    set $E$ of edges, for each clause $c_j$ a complete subgraph of size $K$
    is formed by connecting all pairs of ``literal nodes'' $(kj)$ pairwise
    which correspond to this clause. Also, each ``variable node'' $i$ is
    connected with its corresponding ``negated variable node'' $\overline{i}$.
    Finally, for each literal $l_{kj}$, if the literal represents a non-negated
    variable $x_i$, an edge connecting $(kj)$ with $i$ is included,
    while if the literal represents a negated
    variable $\overline{x}_i$ the corresponding literal node $(kj)$ is connected
    with $\overline{i}$. Thus, $E$ contains $MK(K-1)/2+N+MK$ edges.
    Now a minimum vertex cover is obtained. This is a subset $V'\subset V$ of
    nodes such that for each edge of $E$ at least one of the two endpoints is in
    $V'$. By construction \cite{garey1979}, $G$ contains a vertex cover of size
    $N+(K-1)M$ if and only if the corresponding formula is satisfiable.
    In Fig.~\ref{fig:vc:sat} the graph corresponding to the formula from
    Eq.~\eqref{eq:sat_example} is shown.

    Thus, one approach to SAT is to transform each formula into the equivalent
    graph and use an existing algorithm for VC to solve it. We applied
    an LP formulation with cycle cutting planes, see Ref.~\cite{dewenter2012phase}
    for details. For the previous work, this algorithm
    was able to solve VC instances in the parameter-space region, where the
    solutions were contained basically in one cluster, corresponding to
    the replica-symmetric region \cite{weigt2000number}.

\section*{Results}

    We sample random 3-SAT instances, where each clause may contain any variable
    at most once. For up to $14$ system sizes $N \in [128, 524288]$
    we simulated $n = 5000$ realizations for $30$ to $100$ different values of
    the density $\alpha$. For comparison, we also performed
    simulations for 2-SAT and 4-SAT, with a smaller range of sizes, see
    below. All error
    estimates are obtained by bootstrap resampling \cite{efron1979,young2015,hartmann2015practical},
    except for errors of fit parameters shown in the plots, which are
    \emph{gnuplot}'s asymptotic standard errors corrected according to Ref.~\cite{young2015}.
    To solve the LP realizations, the implementation of the dual-simplex
    algorithms of the commercial optimization library \emph{CPLEX} \cite{cplex} is used.
    During the research additionally the primal- and dual-simplex implementations
    of \emph{Gurobi} \cite{gurobi} and \emph{lp\_solve} \cite{lpsolve} with multiple pricing strategies were
    used to ensure that the results are independent from the algorithm and the
    details of the implementation.

\subsection*{LP-Transitions with Objective Function}
\label{sec:lp}
    As mentioned before, we observed that
    non-trivial objective functions can be used to obtain a
    result independent from the details of the LP-solver implementation.
    Though the objective function itself will have an influence on the position
    of the transition point. An objective which prefers variables to be integer
    will result in more integer solutions at the same value of
    $\alpha$, i.e., yield a
    transition at a larger value of $\alpha$.

    \begin{figure}[htb]
        \center
        \includegraphics[scale=1]{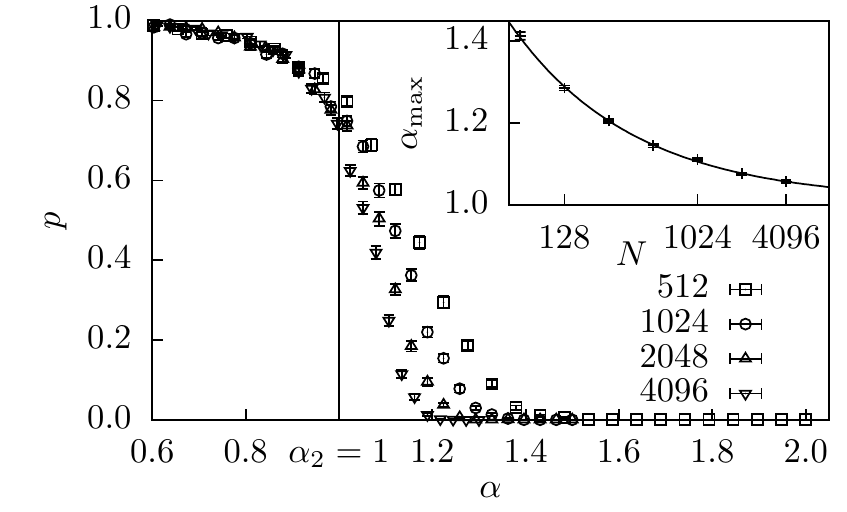}
        \caption{
        \label{fig:twosat}
            Solution probability $p$ that SMM yields an integer solution
            for 2-SAT. The
            inset shows a extrapolation to large $N$ for a transition from
            solvable to unsolvable. The extrapolation is a power law with
            offset $\alpha_{\mathrm{max}} = aN^{-\tilde{b}} + \alpha_{2}$
            and yields a transition of $\alpha_2 = 1.011(9)$ with a goodness of
            fit $\chi_\mathrm{red}^2 = 0.6$.
        }
    \end{figure}

    First, we will demonstrate that this method can be used to
    solve instances of $K$-SAT efficiently for a range
    of values of $\alpha$, and that there exists a phase transition
    to a hard (or unsolvable) region.
    For the $2$-SAT problem, which is not NP-complete,
    it is known that the SAT-UNSAT transition happens at $\alpha_2 = 1$ \cite{Goerdt1996}.
    For $2$-SAT there are also exact polynomial time algorithms. So
    when applying our linear programming approach, we would assume that in the limit
    of large $N$ for $\alpha < 1$, all instances are easily solvable and for
    $\alpha > 1$ not. This test does indeed work out when using the SMM objective function.
    In Fig.~\ref{fig:twosat} the probability that a realization is solvable by
    the LP+SMM approach, i.e., the solution is integer,
    is shown as a function of the clause density $\alpha$. Around
    $\alpha = 1$ the behaviour switches from solvable to not-solvable.
    The decrease in probability to solve is steeper for larger system sizes $N$,
    which is a behaviour typical for phase transitions. The inset
    shows the positions of maximum variance, since $p$ is binomial, this
    corresponds to the positions where half of the realizations are solvable.
    Therefore, this method effectively extrapolates the position of the $p=0.5$
    point, which should be at the transition point for $N\to\infty$.
    We performed the extrapolation
    by using a typical finite-size scaling power-law
    ansatz $\alpha_{\mathrm{max}} = aN^{-\tilde{b}} + \alpha_{2}$,
    like in previous work \cite{dewenter2012phase,schawe2016phase}.
    This yielded a critical point of $\alpha_2 = 1.011(9)$, which is
    in very good agreement with the expectation, especially considering the
    small system sizes used for this extrapolation. Note that
    without the objective function $2$-SAT is also susceptible to the greater
    degeneracy and the above mentioned problems become visible, leading to no
    clear result (not shown here).
    This result for $2$-SAT shows that the method per se is
    a valid approach to our question.

\begin{figure}[htb]
        \subfigure[]
        {
            \label{fig:satLitOpt}
        }
        \subfigure[]
        {
            \label{fig:satLitOptPeaks}
        }
        \includegraphics[scale=1]{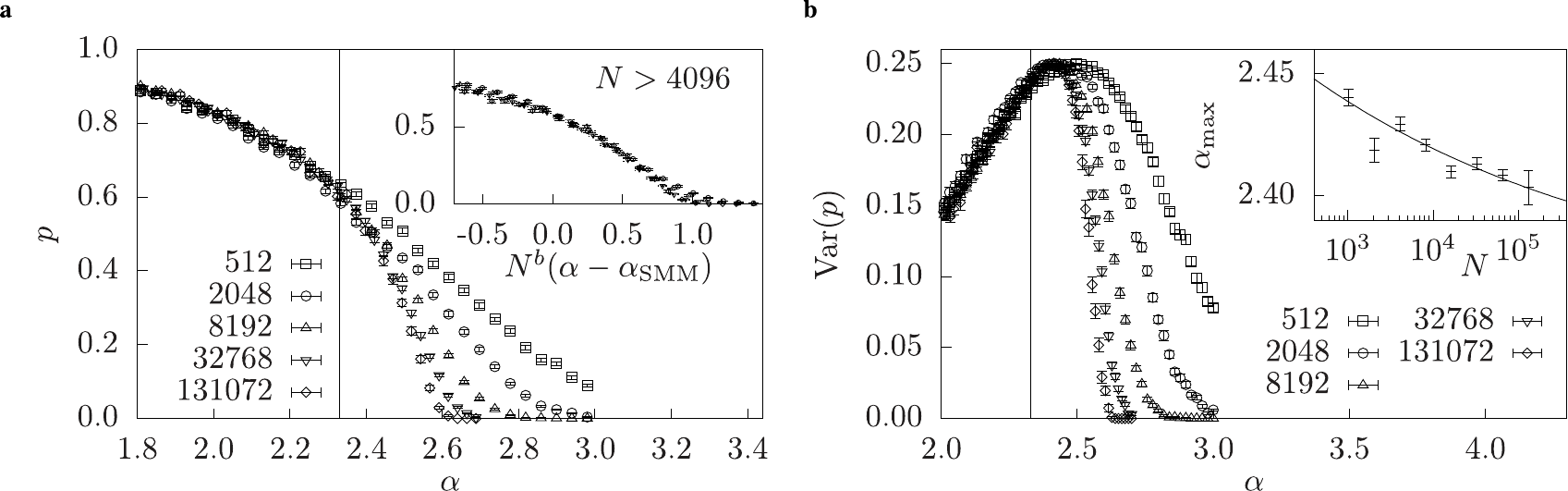}
        \caption{
        \label{fig:sat}
            (\textbf{a}) Probability $p$ that SMM yields an integer solution.
            The smaller system sizes show visible deviations
            from the common curves, which is visible in the main plot, where
            $N=512$ does not lie on the other curves at $\alpha_\mathrm{SMM}$
            marked by the vertical line.
            Inset: Collapse with $b = 0.118(1)$, $\alpha_\mathrm{SMM} = 2.36(1)$
            for $N\ge 4096$.
            (\textbf{b}) Variance of the solution probability $p$ for the SMM objective.
            Inset: The power-law fit $\alpha_{\mathrm{max}} = aN^{-\tilde{b}} + \alpha_{\mathrm{SMM}}$
            to the position of the peaks $\alpha_{\mathrm{max}}$. For fixed
            $\tilde{b} = 0.1176$ the fit yields $\alpha_\mathrm{SMM} = 2.35(1)$
            with $\chi^2_{\mathrm{red}} = 1.9$.
        }
    \end{figure}

    The same procedure is performed for $3$-SAT in Fig.~\ref{fig:sat}.
    The transition depicted here is an algorithmic
    transition from \emph{easy}, since most realizations are solvable by
    LP techniques, i.e., in polynomial time, to some \emph{harder} phase,
    where the LP does not yield solutions.

    To estimate $\alpha_{\mathrm{SMM}}$ for 3-SAT, we use finite-size scaling, i.e.,
    rescaling the $\alpha$ axis according to $(\alpha - \alpha_{\mathrm{SMM}})N^b$,
    should result in a \emph{collapse} of the data points on one curve. To find
    values $\alpha_{\mathrm{SMM}}$ and $b$ generating a good collapse
    the algorithm and implementation from Ref.~\cite{autoscale2009} is
    used. It yields $\alpha_{\mathrm{SMM}} = 2.361(7)$ and $b=0.118(1)$.
    The collapse obtained when rescaling with those values is shown in the
    inset of Fig.~\ref{fig:satLitOpt}. Note that for second-order
    phase transitions \cite{stanley1971} the
    exponent $\nu=1/b$ describes the growth of correlations in the system
    and is one of the \emph{critical exponents}
    used in statistical physics to characterize phase transitions
    according to \emph{universality classes}.

    We could not use the same extrapolation of the peak positions, like for $2-SAT$,
    since the range of $\alpha_\mathrm{max}$ was too small to yield reliable fits.
    But using the fixed exponent from the collapse, leads to a compatible
    critical density $\alpha_\mathrm{SMM} = 2.35(1)$ and a reasonable fit, with
    $\chi_\mathrm{red}^2 = 1.9$, as shown in Fig.~\ref{fig:satLitOptPeaks}.

    Our results for $4$-SAT look similar (not shown) and exhibit an easy-hard
    transition as well. We performed a corresponding analysis. The resulting
    exponent $b$ seems to be larger and a fit through the positions
    of the maxima yields $\alpha_\mathrm{SMM} = 3.60(8)$, $\tilde{b}=0.5(1)$
    with $\chi^{2}_{\mathrm{red}}=0.5$.

    The other optimization function of this study, MV, i.e., maximizing the sum
    of all variables, leads to a qualitatively similar behaviour for 3-SAT
    as SMM but a transition at lower
    $\alpha_{\mathrm{MV}} = 1.25(2)$ (not pictured due to qualitative
    similarity, simulations used smaller system sizes).
    The lower transition point is plausible, since this maximization prefers
    variables to be larger than zero instead of $0$. For this
    reason, we have not analysed this algorithm for 4-SAT.
    The best estimates for the transition points are collected in
    Tab.~\ref{tab:values}.

    \begin{table}[htb]
        \caption{\label{tab:values}
            Values of critical points.
            $\alpha_{\mathrm{VC}}$ denotes the critical point when mapping SAT to VC
            and applying an LP + cutting plane solver used for VC.
            $\alpha_{\mathrm{MV}}$ is the easy-hard transition for LP+MV.
            $\alpha_{\mathrm{SMM}}$ is the easy-hard transition for LP+SMM.
            $\alpha_{\mathrm{c}}$ denotes the clustering transition \cite{krzakala2007gibbs}
            and $\alpha_{\mathrm{s}}$ the SAT-UNSAT transition \cite{mezard2002,mertens2006threshold}.
        }
        \begin{tabular*}{\linewidth}{l @{\extracolsep{\fill}} c c c c c}
            \toprule
                $K$ & $\alpha_{\mathrm{VC}}$ & $\alpha_{\mathrm{MV}}$ & $\alpha_{\mathrm{SMM}}$ & $\alpha_{\mathrm{c}}$ & $\alpha_{\mathrm{s}}$\\
            \midrule
                $2$ &  --       & --        & $1.011(9)$            & --                & $1\phantom{.11}$ \\
                $3$ & $0.90(3)$ & $1.25(2)$ & $2.361(7)$            & $3.86\phantom{1}$ & $4.26$ \\
                $4$ & --        & --        & $3.60(8)\phantom{1}$  & $9.547$           & $9.93$ \\
            \bottomrule
        \end{tabular*}
    \end{table}

    As shortly mentioned above, we also implemented
    cutting planes (CP), which are additional inequalities which do not
    change the nature of any solution but restrict the search space
    of non-feasible, i.e., non-integer configurations.
    In similar studies on VC \cite{dewenter2012phase}
    and the TSP \cite{schawe2016phase},
    the introduction of cutting planes yielded substantially better results.
    For vertex cover the introduction of (potentially
    exponentially many but actually few) CPs
    even lead to an LP+CP transition at the point
    where in the analytic solution replica symmetry breaking, i.e., clustering
    of solutions appeared \cite{weigt2000number}.
    Unfortunately, this efficiency of CP was not observable during our study
    for $K$-SAT. It seems that the cutting
    planes we tried, namely \emph{resolution cuts} \cite{hooker1988resolution}
    and \emph{clique cuts} \cite{atamturk2000conflict},
    were too weak at the low values of $\alpha$ examined here.
    Another cutting plane for the SAT problem, the \emph{odd cycle inequalities}
    \cite{joy1997branch} are not directly applicable for $K$-SAT with
    $K \ge 3$, since they need clauses with 2 variables to be constructed.
    While they are useful as local cuts in a branch and cut procedure,
    they are never violated in the beginning for $K \ge 3$ and thus not
    applicable for this study.

    \begin{figure}[htb]
        \center
        \includegraphics[scale=1]{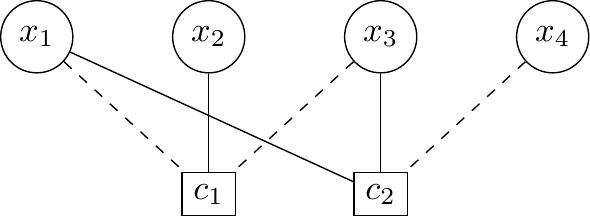}
        \caption{
        \label{fig:fg}
            Factor graph representation of the example Eq.~\eqref{eq:sat_example}.
            Dashed lines represent negative weights, corresponding to negated
            variables.
        }
    \end{figure}

    Next, we investigate whether the observed easy-hard
    transitions correspond to changes of the structure of the
    problem instances, as it was previously found for the
    vertex-cover problem \cite{vertexCover1,vertexCover2,cover-time2001,dewenter2012phase}.
    While VC is defined on graphs anyway, for $K$-SAT
    we study the related graph representation as \emph{factor graph} \cite{Montanari2004instability,mezard2009}.
    This representation is especially useful for \emph{belief} or \emph{survey propagation}
    approaches \cite{Mezard2002Random,braunstein2005survey} but also useful
    to study structural properties.
    In the FG, which is a bipartite graph exhibiting a node for each variable
    and a node for each clause, i.e., containing $N+M$ nodes, each variable is
    connected to the clauses it occurs in with weights representing whether they
    are negated or not. The corresponding graph of example Eq.~\eqref{eq:sat_example}
    is shown in Fig.~\ref{fig:fg}. Note that this representation, when
    disregarding the weights, is equivalent to a hypergraph representation,
    where every clause is represented by a hyperedge connecting $K$ variables.

    We looked at some known points of the factor graph, where
    its properties change, which could plausibly influence the hardness of the
    problem. Note that for the NP-complete VC, it was
    possible to relate the percolation threshold and the threshold,
    beyond which the leaf removal heuristic~\cite{bauer2001core} does not
    yield solutions anymore (which coincides with the appearance of
    replica-symmetry breaking and solution-space clustering),
    to the points where the problems turns harder also for
    special formulations of the linear programming approach
    \cite{dewenter2012phase}.
    We will list the properties for $3$-SAT we looked at. Most
    transitions are known from the literature, but few not, to our knowledge.
    For those we numerically investigated the transitions during this study
    and present the result shortly here,
    which also contributes to the characterization of the 3-SAT ensemble.

    \begin{itemize}
        \item The \emph{percolation threshold}, i.e., the point below which
        there will be no connected component of size $\mathcal{O}(N)$ is
        at $\alpha = 1/6$ \cite{mezard2009}.
        This means that below this threshold clauses do typically not
        share variables and are therefore largely independent, i.e., it
        should be rather easy to solve for almost any algorithm.
        \item The transition where the remaining \emph{pure literal core} is of order
        $\mathcal{O}(N)$, i.e., the value of $\alpha$, beyond which the pure
        literal rule, which is an integral ingredient for the classical DPLL
        search, does not lead to solutions anymore is at
        $\alpha_\mathrm{pl} = 1.636..$ \cite{molloy2005cores}.
        The pure literal rule is to remove pure variables and their clauses from
        the problem. Variables are \emph{pure}, if they are appearing only in one
        polarity and can therefore be always set to fulfil all their clauses.
        Note that for 2-SAT the pure literal rule works up to
        the SAT-UNSAT threshold $\alpha_2 = 1$~\cite{molloy2005cores},
        coinciding with the solvability transition of SMM.
        \item The \emph{unit clause rule} shows its transition above the
        LP+SMM case at $\alpha_\mathrm{uc} = 8/3 \approx 2.66$~\cite{achlioptas2001lower}.
        \item For a naive \emph{$q$-core} analysis, where the core is the set
        of remaining nodes after all nodes with degree lower than $q$ are
        iteratively removed, we treated the clause nodes the same as
        the variable nodes. It yielded a percolation transition,
        i.e., the existence of a $q$-core of order of graph size at
        $\alpha_\text{2-core} = 0.223(3)$ and
        $\alpha_\text{3-core} = 1.554(1)$ from our measurements.
        \item The \emph{$q$-core} transitions for a hypergraph approximation of
        $K$-SAT, where $K$ nodes are connected by single hyperedges, are
        known exactly. Note that the $2$-core for this ensemble is equivalent
        to the pure literal rule according to~\cite{molloy2005cores}. The
        appearance of a three core occurs at $\alpha_\text{3-CORE} = 4.2847..$~\cite{molloy2005cores}.
        \item The \emph{leaf removal rule}, which is a valid heuristic for the related
        XOR-SAT problem and the vertex cover problem, is known to have a
        transition at $\alpha_\mathrm{lr} = 0.81847..$\cite{Mezard2003}.
        Interestingly, although not obviously related, this is half of
        $\alpha_\mathrm{pl}$ of the pure literal rule.
        \item The transition where a \emph{biconnected component}~\cite{Hopcroft1973algorithm,cormen2009introduction}
        appears, i.e., a connected component, in which every pair of nodes
        stays in the same connected component if any other node is removed,
        happens at $\alpha_\mathrm{bi} = 0.190(20)$.
        Similarly, the \emph{bi-edge-connected component}, from which an
        arbitrary edge can be removed while staying connected, shows the transition
        at $\alpha_\text{bi-edge} = 0.211(7)$.
        This value is close to the appearance of our naive $2$-core,
        which is plausible since a $2$-core consists of biconnected components
        possibly connected by single edges.
    \end{itemize}

\begin{figure}[htb]
        \center
        \includegraphics[scale=1]{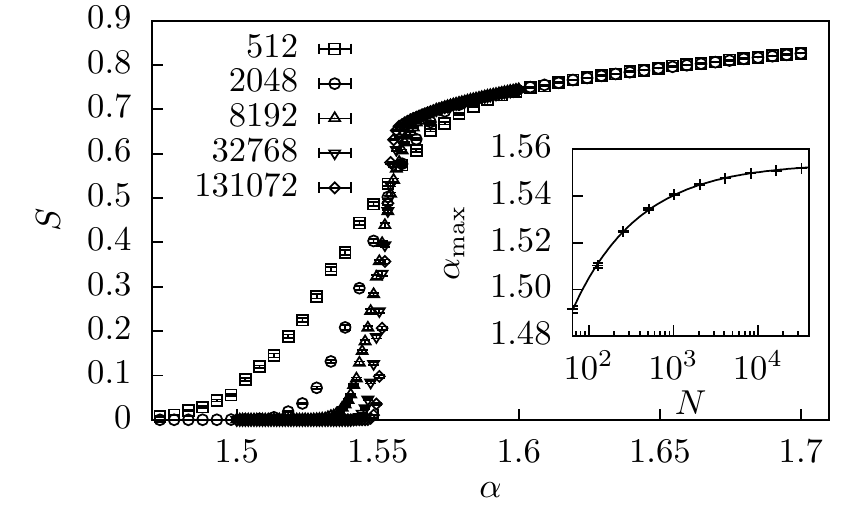}
        \caption{
        \label{fig:3core}
            Relative size of the naive $3$-core of the factor graph at different
            values of $\alpha$. The inset shows the positions of the maxima of
            the peaks of the variance and an extrapolation according to a power
            law with offset, resulting in the above listed values.
        }
    \end{figure}

    The process to determine the transition points for the cases, where we did
    not find literature values, is very similar to the process shown in Fig.~\ref{fig:twosat}.
    A example for the naive $3$-core on the factor graph is shown in Fig.~\ref{fig:3core}.
    Again we extrapolated the position of the peaks of the variance to large $N$
    using a power law with offset $\alpha_{\mathrm{max}} = aN^{-\tilde{b}} + \alpha_{\text{3-core}}$.

    When comparing the values of these transitions to the easy-hard
    transitions points listed in table~\ref{tab:values}, one observes no
    coincidence. Thus, in contrast to the previously studied VC,
    there exist no coincidence with a ``simple'' change of the problem
    structure, which might serve as an explanation of the observed easy-hard
    transitions.

    Finally, the existence of correspondences between easy-hard
    transitions and structural percolation transitions for the vertex-cover
    problem motivated us to perform the following test:
    we also used for $K=3$ the mapping of $K$-SAT
    to VC for formulas up to $N=10000$.
    Using LP and cycle cutting planes \cite{dewenter2012phase} the
    corresponding equivalent instances VC were
    solved for various values of $\alpha$.
    Again we measured the probability that an instance was solved
    by an integer solution as a function of $\alpha$, for different system
    sizes. Using an analysis (not shown) as for the previous approaches,
    we were able to extrapolate an easy-hard transition for this point.
    We obtained a critical value of $\alpha_{\mathrm{VC}}=0.90(3)$, which
    is well below the easy-hard transitions obtained using the LP-based
    approaches presented above.
    Therefore, apparently it does not pay off using a mapping
    to another problem, at least for this pair of problems. A mapping to TSP
    will lead to TSP instances which are quite large for decent SAT realizations,
    therefore we did not perform simulations for this.

    It is rather intriguing that SAT behaves so differently in comparison to
    vertex cover, despite their very close relation. While we can not give
    the reason for this behaviour, we not only have identified
    several easy-hard transitions in this study, but have also shed
    some light on this apparently
    fundamental difference. In the future, maybe other approaches
    can yield a better understanding of this perplexing fact.

\section*{Conclusions}
\label{sec:conclusion}
    We study the solvability of
    random $K$-SAT realizations at different values of the
    clause-to-variable density $\alpha$ using linear programming.
    A realization is solved if the LP yields an integer solution.
    Since there are LP algorithms that run in polynomial time, this means
    such a realization is ``easy'' in regard to this algorithm.

    Since SAT is a decision problem, the application of pure LP actually does
    not involve optimization, which seems to make the problem less well-behaved.
    This is probably the reason that also for practical applications
    LP has not been used widely for SAT and instead methods like the
    branch-and-bound approach DPLL dominate. Nevertheless,
    the investigation of the behaviour of any algorithm, even not the most
    efficient one, might lead to
    a better understanding of the specific problem structure and of the sources
    of computational hardness in general. To allow for an investigation
    of SAT using LP, we included artificial objective
    functions, which makes the problem numerically better behaved.
    Note that we also observed that usually more realizations can be
    solved when including an artificial objective function as compared
    to the trivial objective function.

    We were able to identify several easy-hard transitions,
    depending on the algorithm which was used. The
    inclusion of artificial objective functions lead to higher values
    of the control parameter $\alpha$, where solutions can still be found.
    In contrast to previous work on VC on random graphs, none of these transitions
    coincided with the onset of a clustered solution landscape.
    The reason  might be that $K$-SAT, or ``natural''
    decision problems in general behave differently. On the other hand,
    random $K$-SAT exhibits several transitions for the structure of the
    solution landscape anyway, in contrast to VC on random graphs. It would be interesting
    to investigate the relation between LP and the solution landscape more
    thoroughly in the future.

    Anyway, from the practical point of view,
    we wonder if carefully crafted objective functions
    could be used to improve the efficiency of solving decision problems
    with an LP approach, such as branch and cut.

    Like for the clustering transitions, we were not able
    to identify other structural transitions
    coinciding with an observed easy-hard transition
    detected by any of the tested LP formulations. Nevertheless, we believe
    that such a coincidence should exist, since the properties of a problem
    should be coded in the graph structure. Therefore, this unknown
    structural property seems to be of a more complex type.
    The fact that the relation between LP-based easy-hard transitions
    and structural properties appears to be more complex might
    also be related to the mentioned richer behaviour of the solution landscape as
    a function of the density $\alpha$ (in comparison with VC).
    Thus, here is still some work to be done.

    Furthermore, in contrast to
    the previously studied vertex cover problem and the travelling salesperson
    problem, we did not observe any improvement by applying cutting planes.
    This could be due to the type of cutting planes used. Finally,
    we applied a mapping of SAT to VC and used specific VC LP+cutting plane
    algorithm, which is able to solve standard Erd\H{o}s-R\'enyi
    VC instances just up to the clustering threshold.
    Nevertheless, for SAT this did not pay off, the easy-hard transition
    appears to be at even smaller values of $\alpha$ compared to the MV
    objective.

    \bibliography{lit}

    \section*{Acknowledgements}
        This work was supported by the German Science Foundation (DFG) through
        the grant HA 3169/8-1.
        The simulations were performed at the HPC Cluster HERO and CARL, both
        located at the University of Oldenburg (Germany) and funded by the DFG
        through its Major Research Instrumentation Programme
        (INST 184/108-1 FUGG and INST 184/157-1 FUGG) and the Ministry of
        Science and Culture (MWK) of the Lower Saxony State.

    \section*{Author contributions statement}
        A.K.H. conceived the project, R.B. and H.S. developed the simulation.
        All three authors analysed the results and contributed to the manuscript.

    \section*{Additional information}
        \textbf{Competing financial interests}
            The authors declare no competing financial interests.

\end{document}